# An Automated FPGA-based Framework for Rapid Prototyping of Nonbinary LDPC Codes

Yaoyu Tao[1], Qi Wu[2]

[1]Department of Electrical Engineering,
Stanford University, CA, USA

[2]University of California,
Davis, CA, USA

*Abstract*— **Nonbinary LDPC codes have shown superior performance close to the Shannon limit. Compared to binary LDPC codes of similar lengths, they can reach orders of magnitudes lower error rate. However, multitude of design freedoms of nonbinary LDPC codes complicates the practical code and decoder design process. Fast simulations are critically important to evaluate the pros and cons. Rapid prototyping on FPGA is attractive but takes significant design efforts due to its high design complexity. We propose a high-throughput reconfigurable hardware emulation architecture with decoder and peripheral co-design. The architecture enables a library and script-based framework that automates the construction of FPGA emulations. Code and decoder design parameters are programmed either during run time or by script in design time. We demonstrate the capability of the framework in evaluating practical code and decoder design by experimenting with two popular nonbinary LDPC codes, regular (2, $d_c$) codes and quasi-cyclic codes: each emulation model can be auto-constructed within hours and the decoder delivers excellent error-correcting performance on a Xilinx Virtex-5 FPGA with throughput of up to hundreds of Mbps.**

## I. INTRODUCTION

Nonbinary low-density parity-check (NB-LDPC) codes designed in high order Galois fields have shown great potential approaching the Shannon limit [1], [2]. However, practical performance of nonbinary LDPC codes can be far from their theoretical performance, sometimes even worse than binary LDPC codes, for the following two reasons: 1) practical decoding algorithms, like extended min-sum (EMS) [3] and min-max (MM) [4] with message truncation [5] or skimming [6], introduce performance degradations and the effect of high order Galois field may be diminished; 2) practical decoder implementation inject non-idealities, such as finite word length and fix-point quantization effects [7]. Therefore, it is critically important to evaluate code and decoder design for each new application that is brought into consideration to ensure the efficiency gain by deploying NB-LDPC codes.

FPGA emulations are widely used to accelerate the simulations showing orders of magnitude speedup. A typical simulation setup for NB-LDPC contains a decoder under study and peripherals such as source generator and channel model. Dedicated emulations have been developed recently for specific NB-LDPC codes to enhance the simulation throughput from less than a hundred kb/s on a microprocessor up to hundreds of Mb/s. A high-speed non-binary LDPC decoder based on trellis MM algorithm with layered schedule achieved 630Mbps for a (2304, 2048) NB-LDPC code over GF(16) in [8]. Prior work [6] also demonstrated FPGA emulation for (960, 480) regular-(2, 4) codes delivering 9.76 Mb/s throughput based on four parallel EMS decoders with message truncation and skimming. The (744, 653) quasi-cyclic code over GF(32) enabled a partial parallel MM decoder that achieved a 9.3Mb/s throughput [9], [10]. However, designing FPGA emulation is not as easy as writing software codes and it's especially difficult to implement reconfigurable property on hardware as software parameters. Besides, it also requires extensive efforts in creating dedicated hardware architecture and running through FPGA synthesis. For high complexity designs like NB-LDPC, it often takes weeks to months to get a working FPGA emulation model. These barriers render rapid FPGA prototyping inaccessible to NB-LDPC which would otherwise benefit significantly in code and decoder evaluation.

In this work, we focus on creating a FPGA-based framework enabling end-to-end automated design starting from code and decoder design parameters to a working FPGA emulation. High-throughput reconfigurable hardware emulation architecture incorporating EMS and MM decoder is proposed to address the challenges in creating FPGA emulation. The framework is built upon a library that consists of elementary building blocks, and a set of router scripts that assemble the building blocks to a complete emulation model. We also demonstrate the capability of the framework, especially in low error rate regime, by experimenting with two popular NB-LDPC codes that have been considered for practical adoptions: regular-(2, $d_c$) codes [11] and quasi-cyclic codes [12]. In all cases the framework is able to complete decoder construction within hours including FPGA synthesis. The resulting emulation models are tested on Virtex-5 FPGA, delivering throughputs up to hundreds of Mbps and reaching a BER of $10^{-9}$ in a day.

## II. BACKGROUND

NB-LDPC code is defined by a parity-check matrix H of size $m \times n$, where $n$ is the block length and $m$ is the number of parity checks. The elements in the H matrix belong to Galois field GF($q$) [1], [2]. The H matrix can also be represented by a factor graph, where each column is mapped to a variable node (VN), each row to a check node (CN), and an edge connects variable node $v_j$ and check node $c_i$ if H($i, j$) ≠ 0. A regular ($d_v$, $d_c$) NB-LDPC code has constant column weight $d_v$ and row weight $d_c$. We surveyed in the literature and summarized that almost all the NB-LDPC codes considered for practical implementation have regular H matrices [12], [13], [14], [15] due to its significantly lower hardware complexity.

Several efficient algorithms and their variations have been proposed for NB-LDPC with various error-correcting performance and implementation complexity [1]-[7], [13]. Among them the EMS [3] and the MM [4] algorithms work remarkably well for hardware decoder design: they achieve a performance close to the original BP algorithm [1], [2] and their complexity is relatively low. Message truncation [5], bubble-check [16] and skimming [6] techniques claim even lower complexity and demonstrates great potential for practical, but they also degrade the error correction performance.

TABLE I NB-LDPC FPGA EMULATION PARAMETERS

| Category | Parameters | Description |
|---|---|---|
| Code parameters | $q$ | GF field order |
| | $m$ | Number of rows in H matrix |
| | $n$ | Number of columns in H matrix |
| | $d_v$ | Column degree or variable degree |
| | $d_c$ | Row degree or check degree |
| | $p$ | Non-zero positions in H matrix |
| | $e$ | GF indices in H matrix |
| Decoder design parameters | $n_m$ | Message truncation number |
| | $Q$ | Number of quantization bits |
| | $L_{S\text{-}VN}$ | VN sorter length |
| | $L_{S\text{-}CN}$ | CN sorter length |
| Run-time parameters | $L$ | Iteration limit |
| | $F$ | Frame limit |
| | $SNR$ | Signal-to-noise ratio in dB |

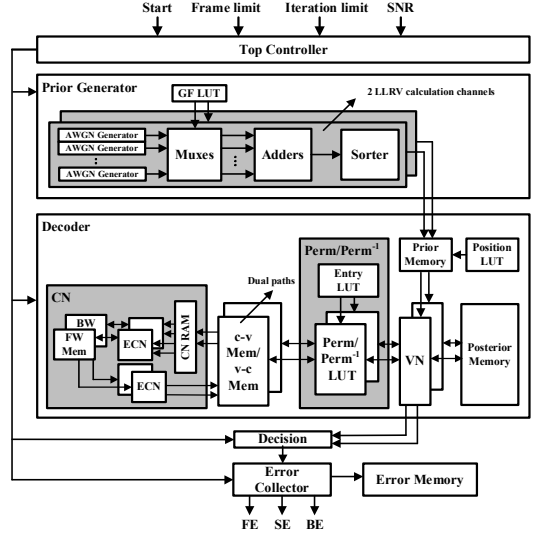

Fig. 2. Reconfigurable emulation system architecture

We briefly summarize the EMS and MM algorithms here for completeness. Both algorithms follow a five-step decoding process as follows: (1) each variable node is initialized with sorted *prior* log-likelihood ratio vectors (LLRV) L along with their associated GF indices $\beta_L$. The length of LLRVs is determined by message truncation number $n_m$; (2) variable-to-check (v-c) messages are permuted based on the H matrix and sent to the check nodes. In the first iteration, the *priors* are used as the v-c messages; (3) for each adjacent variable node $v_j$, check node $c_i$ computes the check-to-variable (c-v) message $\{V_{ij}[k]\}, k \in \{0, \ldots, n_m - 1\}$, that the parity-check equation is satisfied if $v_j = \beta_{V_{ij}}[k]$. The computation is implemented as a *forward-backward recursion*: EMS computes the sums through this recursion while MM picks only the maxes without summation operations for even lower complexity. Note that c-v messages are sorted and only the $n_m$ highest probabilities are stored in both algorithms. Bubble-check technique [16] improves the check node latency as well as hardware utilization by reducing the sorter length from $n_m$ to $L_{S\text{-}CN}$ while still maintaining the equivalent functionality; (4) c-v messages are inverse permuted before being sent to the variable nodes; (5) variable node $v_j$ computes the v-c message $\{U_{ji}[k]\}, k \in \{0, \ldots, n_m - 1\}$ for each adjacent check node $c_i$ based on the *prior* LLRVs and the permuted c-v messages. Skimming technique [6] skims less reliable probabilities and reduces VN sorter length from $n_m$ to $L_{S\text{-}VN}$. The procedure repeats itself from step (2) until iteration limit $L$.

Decoder architectures implementing EMS or MM have been developed for FPGA emulation of various NB-LDPC codes [6], [8], [9], [10], and most of them have limited flexibility for parameters like iteration limit $L$; however, important parameters that are significant factors of error-correction performance and throughput, like $q$ and $n_m$, can only be studied in software simulations which take weeks to months to reach low BER region. Reconfigurability for these important parameters on hardware involves complicated architecture and schedule changes and takes extensive efforts and time repeatedly for every possible parameter combination. The challenges call for an automated design flow with new decoder and emulation architecture that enables full reconfigurability and delivers a high throughput.

III. RECONFIGURABILE EMULATION

Reconfigurable emulation for NB-LDPC requires addressing parameters of three categories: code parameters, decoder design parameters and run-time parameters. We summarize the parameters with their descriptions as in Table I.

*A. Emulation System Design*

Suppose without loss of generality that all-zero codeword are transmitted, we introduce a fully reconfigurable emulation system with high-throughput decoder.

Fig. 2 shows the architecture of proposed emulation system. A top controller implementing a finite state machine orchestrates the emulation. The emulation system stays at IDLE state until input Start jumps from 0 to 1. System then enters the RUN state. There are two sub-states in the RUN state: (1) Prior Generation (PG), (2) Decode and Decision (DD). They iterate for each frame and a counter COUNT keeps track of the RUN state and increments by 1 every time the system reaches the end of DD state. The state transition diagram is shown in Fig. 3.

In PG state, LLRV calculation channels in the prior generator compute sorted LLRVs of length $n_m$ along with their corresponding GF indices and store them to a dual-port prior memory. In each channel, $\log_2(q)$ parallel AWGN generators produce $\log_2(q)$ parallel LLRs of $Q$-bit and send them to a multiplexer array. The multiplexer array also reads $\log_2(q)$ bits that represent a GF($q$) symbol from the GF LUT: Note that each bit is associated with a LLR. The multiplexer array selects LLR if the associated bit is 1 and passes a 0 if the associated bit is 0. A $\log_2(q)$-input adder sums up the outputs from multiplexer array for the symbol LLR and send the result to a sorter of length $q$. It takes $q$ cycles to complete the sorting and another $n_m$ cycles to complete LLRV writes into the prior memory. Two channels are instantiated in our design to make full use of the two ports on prior memory. Assume the latency overhead from AWGN generators to the input of the sorter is $T_{overhead}$, it takes $T_{overhead}+(n/2)\times(q+n_m)$ cycles to complete the entire prior memory initialization.

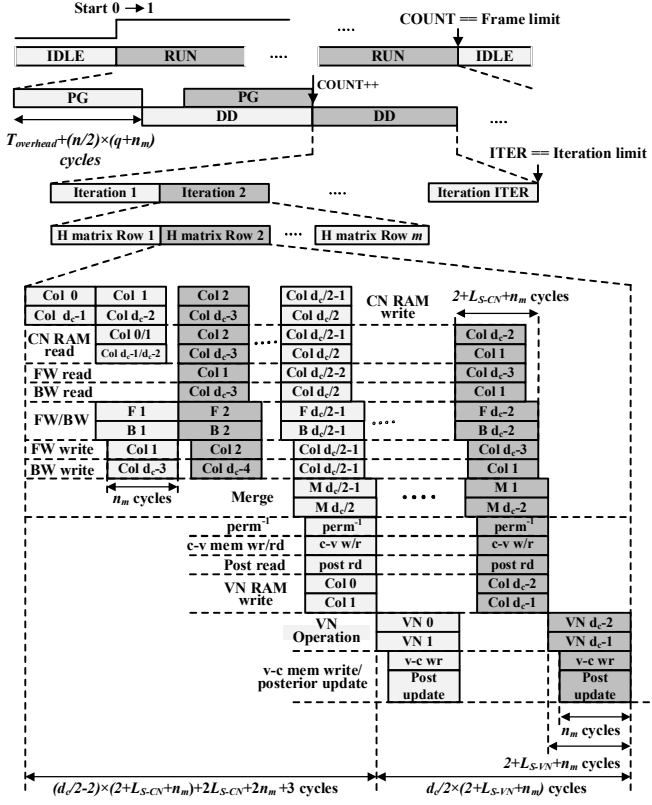

Fig. 3. Scheduling of the reconfigurable emulation system

The EMS or MM decoder is the most complex block of the decoder emulation model. Upon completion of PG state, decoder is fired up for decoding and the system enters the DD state. The DD state consists of ITER number of iterations, in which priors are decoded following a layered row-by-row manner as shown in Fig. 3.

For the first iteration, VNs read the prior LLRVs according to the position LUT and by-pass them to the permutation blocks. The permutation blocks perform GF multiplications and divisions based on the permutation LUT and output permuted v-c messages into v-c memory of size $m\times d_c\times(Q+q)$ bits. A $d_c$-banked dual-port RAM of size $(Q+q)\times n_m$ bits each is used as a buffer and provide required memory access bandwidth for an optimal pipeline schedule like in Fig. 3. For both EMS and MM decoder, CN implements the forward-backward recursion on a $d_c$-stage trellis in three elementary steps: (1) forward step (F), (2) backward step (B), and (3) merge step (M). Each step in the recursion is done by an elementary CN (ECN) with sorter of length $L_{S-CN}$. Note that maximum number of elementary steps that can run simultaneously is 4. Hence, we designed a CN with 4 ECNs and a pair of forward and backward memory that store intermediate messages on trellis. Two ECNs perform the forward and backward, respectively, and each elementary step takes $2+L_{S-CN}+n_m$ cycles to complete. When forward and backward reach the middle of the trellis at $d_c/2$, the other two ECNs start the merge operations and send out two c-v messages to the inverse-permutation blocks.

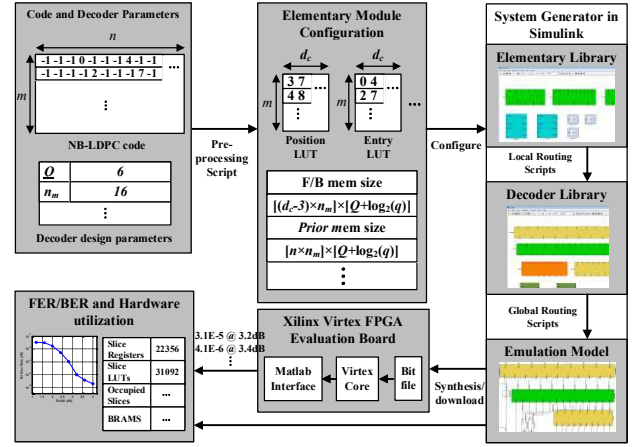

Fig. 4. Automated design flow based on Xilinx FPGA Platform

Inverse permuted c-v messages are then stored into the c-v memory of size $m\times d_c\times(Q+q)$ bits and at the same time written to the VN RAMs. VNs are implemented similar to [6] and they wait until the completion of VN RAM write since they require full-length LLRVs for address look-up, and the latency of each VN operation is $2+L_{S-VN}+n_m$ cycles. The latency of decoding a complete row is shown in Fig. 3 and the next iteration starts by reading v-c memory upon completion of all rows in previous iteration. Note that posterior memory is completely updated only at the end of iteration until all neighboring c-v messages are incorporated for each VN. Decisions are made concurrently with the last-row posterior memory write. The decoder and peripheral co-design enables perfect interleaving of the dual-path with an optimal pipeline schedule shown in Fig. 3.

## IV. AUTOMATED FLOW FOR RAPID PROTOTYPING

We developed an end-to-end automated design flow based on proposed reconfigurable emulation architecture targeting Xilinx Virtex FPGA platform. The flow involves three automation steps: (1) pre-processing, (2) decoder library generation, and (3) top-level routing.

Prior to the automation steps, a Simulink elementary library has been developed that consists of elementary blocks that make up an emulation system: ECN, VN and LUTs, etc. The elementary blocks are designed using Xilinx blockset that can be readily synthesized and they are quick to design and easily reusable. Note that controllers are designed with Mcode blocks that can be programed directly by finite-state-machine written in Matlab codes.

The first step involves a pre-processing Matlab script that computes the configuration parameters referenced in the elementary library. It also picks the ECN and VN with sorter length $L_{S-CN}$ and $L_{S-VN}$, respectively, and creates a list of elementary blocks that are required for given NB-LDPC code and decoder parameters.

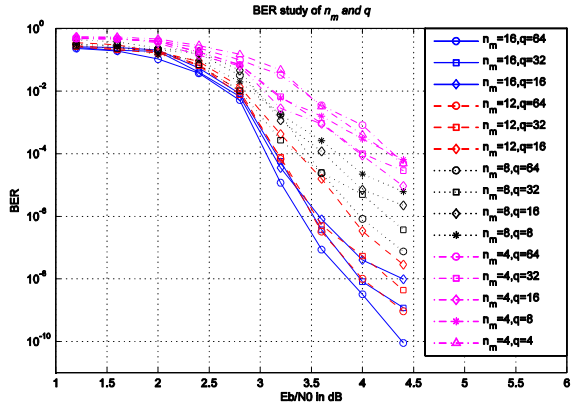

Fig. 5. Performance of rate-1/2 960-bit (2,4)-regular NB-LDPC codes with 6-bit EMS decoder

In the second step, a set of local routers connect elementary blocks together and produce larger building blocks for top level routing. Complete prior generator, CN and Perm/Perm$^{-1}$ processing blocks are obtained, and a decoder library is established that consists of modules ready for top-level routing.

Upon completion of all the modules in decoder library, the top-level router connects them into a complete emulation model in the third step. Interfaces are also added to provide run-time parameters and capture FER/BER. The three-step automation is followed by synthesis that generates bit file, which takes up to several hours based on the experiments we carried out. Excluding the initial efforts in making the reusable elementary library, the proposed framework including synthesis completes within hours.

## V. Experiments and Analysis

We show the capability of proposed automated framework by experiments with two popular NB-LDPC codes considered for hardware implementations, regular $(2, d_c)$ codes and quasi-cyclic (QC) codes.

*Experiment I*: Suppose we are aiming to design a 6-bit decoder based on a rate-1/2 960-bit (2, 4)-regular NB-LDPC [11] targeting BER $10^{-6}$ at SNR 4.4dB with 10 decoding iterations. We studied the BER performance by varying two important parameters, $q$ and $n_m$, based on EMS algorithms without message skimming. With proposed framework, we run emulations on Xilinx Virtex-5 FPGA core at a clock frequency of 120 MHz sweeping $n_m$ from 4 to 16 for various $q$. Note that varying $q$ only changes the values of non-zero entries but doesn't affect the positions of non-zero entries in the H matrix. Fig. 5 shows the BER vs SNR under each possible ($n_m$, $q$) combinations. As discussed in Section III, decoding latency is proportional to $n_m$. Hence a smaller $n_m$ is desired for higher throughput upon meeting the BER spec. Table II shows the hardware utilization for $n_m > 8$ with various $q$ values. Bigger $q$ and $n_m$ result in larger hardware utilization. A combination of $n_m = 8$ and $q = 32$ gives the best throughput and area for the given BER spec.

*Experiment II*: We also experimented our framework with 1024-bit GF(16) QC NB-LDPC codes [12] with code rate $R$

TABLE II FPGA Mapping Results
(based on Xilinx Virtex-5 XC5VLX155T)

| Resource | $n_m = 8$ $q = 32$ | $n_m = 8$ $q = 64$ | $n_m = 12$ $q = 16$ |
|---|---|---|---|
| Slice Registers | 13,929 (15%) | 16742 (18%) | 16,046 (17%) |
| Slice LUTs | 17,210 (17%) | 19,511 (19%) | 17,908 (17%) |
| Occupied Slices | 6,832 (28%) | 8,744 (36%) | 7,167 (30%) |
| BRAMs | 55 (26%) | 59 (28%) | 53 (25%) |

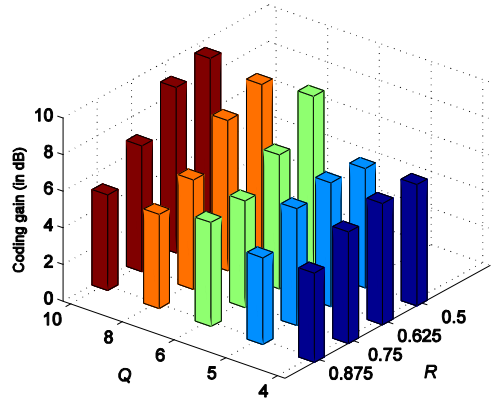

Fig. 6. Coding gain of 1024-bit QC-regular NB-LDPC codes with $Q$-bit MM decoder and $n_m = 16$

varying from 1/2 to 7/8. Note that for QC codes we can maintain regularity of the H matrix if we add or remove a complete row of sub-matrices; hence the code rate $R$ is determined by the number of rows $n$ in the H matrix. Parameter $d_v$ also changes with varying $R$. Fig. 6 shows the coding gain for each possible ($R$, $Q$) combinations based on a $Q$-bit MM decoder with $n_m = 16$. With this study decoder designers can easily pick best decoder configurations for desired coding gain.

## VI. Conclusion

We present a FPGA-based framework with an automated design flow for rapid prototyping of NB-LDPC codes. To the best of our knowledge this work is the first automated FPGA emulation framework for NB-LDPC codes enabling full reconfigurability for code and decoder design parameters. Co-designed emulation architecture with EMS/MM decoder and peripherals is proposed. The framework accepts parameter specifications and produces a complete FPGA emulation model. Experiments on Xilinx Virtex-5 FPGA demonstrate the capability of the framework in evaluating practical NB-LDPC code and decoder design. With parallel copies of decoders mapped onto this FPGA device, the platform delivers excellent error-correction performance with throughput up to hundreds of Mb/s.